\documentclass[journal]{IEEEtran}
\usepackage{times}
\usepackage{graphicx}
\usepackage{amsmath}
\usepackage{bm}
\usepackage{latexsym}
\usepackage{color}
\usepackage{cite}
\begin{document}

\title{Communication Links for Distributed Quantum Computation}

\author{Rodney~Van~Meter,~\IEEEmembership{Member,~IEEE,} Kae~Nemoto,
and~W.J.~Munro \thanks{R. Van Meter is with National Institute of
Informatics, Tokyo, Japan, and with CREST-JST.  K. Nemoto is with
National Institute of Informatics, Tokyo, Japan.  W.J. Munro is with
HP Labs, Bristol, UK.}}

\maketitle

\begin{abstract}
Distributed quantum computation requires quantum operations that act
over a distance on error-correction encoded states of logical qubits,
such as the transfer of qubits via teleportation.  We evaluate the
performance of several quantum error correction codes, and find that
teleportation failure rates of one percent or more are tolerable when
two levels of the [[23,1,7]] code are used.  We present an analysis of
performing quantum error correction (QEC) on QEC-encoded states that
span two quantum computers, including the creation of distributed
logical zeroes.  The transfer of the individual qubits of a logical
state may be multiplexed in time or space, moving serially across a
single link, or in parallel across multiple links.  We show that the
performance and reliability penalty for using serial links is small
for a broad range of physical parameters, making serial links
preferable for a large, distributed quantum multicomputer when
engineering difficulties are considered.  Such a multicomputer will be
able to factor a 1,024-bit number using Shor's algorithm with a high
probability of success.
\end{abstract}

\IEEEpeerreviewmaketitle

\section{Introduction}

\PARstart{D}{istributed} quantum computation uses the physical
resources of two or more quantum computers to solve a single
problem~\cite{grover97:_quant_telec,cleve1997sqe,cirac97:_distr_quant_comput_noisy_chann,dhondt05:_dist-qc,oi06:_dist-ion-trap-qec,steane:ion-atom-light}.
These computers may be geographically distributed, or may be
colocated, with the distributed nature of the system used to overcome
the inherent limitations on the size of a single quantum
computer~\cite{van-meter:qarch-impli,van-meter05:_distr_arith_quant_multic}.
Distributed quantum computation naturally depends on the development
of quantum networking technology to connect the
computers~\cite{elliott:qkd-net,felinto06:_condit}.

A {\em quantum computer} is a device that uses non-classical, quantum
behavior of some physical phenomena to calculate certain functions
asymptotically faster than a purely classical machine
can~\cite{nielsen-chuang:qci,spiller:qip-intro-cp}.  The fundamental
unit of data in a quantum computer is a {\em qubit}, which has two
possible states, written $|0\rangle$ and $|1\rangle$, analogous to the
0 and 1 of a classical bit.  These states may be the horizontal and
vertical polarization of a photon, the up and down spin of a single
electron, or the direction of a single quantum of magnetic flux;
dozens of quantum phenomena have been proposed as qubits, and many of
them are under experimental
evaluation~\cite{van-meter:qarch-impli,spiller:qip-intro-cp,nielsen-chuang:qci}.
Most systems, with the obvious exception of photons, hold qubits in a
register, and execute ``gates'' on the qubits, manipulating their
state like instructions in a classical computer manipulate the bits of
a register.

Perhaps the three most famous quantum algorithms are Shor's algorithm
for factoring large numbers, Grover's search algorithm, and the
Deutsch-Jozsa algorithm for distinguishing among certain classes of
functions~\cite{shor:factor,grover96,deutsch-jozsa92}.  Shor's
algorithm appears to offer a superpolynomial speedup for factoring,
compared to the best known classical algorithm.  Grover has shown
that, for unstructured search problems, the best a quantum computer
can do is to search all $N$ possible solutions in $O(\sqrt{N})$
operations, while Deutsch-Jozsa turns a probabilistic problem into one
with a deterministic, certain answer after a single iteration.  All
three have been demonstrated experimentally at very small
scales~\cite{chuang98:_dj-exper,chuang98:_grover_impl,vandersypen:shor-experiment}.
However, designing and building quantum computers capable of solving
problems at scales that are classically intractable will require many
more years of effort from physicists working on the basic
technologies, theorists designing algorithms including quantum error
correction, and quantum computer architects.

Quantum computation utilizes the quantum characteristics of {\em
superposition}, {\em entanglement}, {\em quantum interference}, and
{\em measurement} to achieve its speedup in computational class.
Superposition, entanglement and interference refer to the wavelike
behavior of a quantum system.  For our qubit, we have two basis
states, $|0\rangle$ and $|1\rangle$, which can be distinguished by
measurement in the computational basis, giving a classical value. A
superposition state contains amplitudes for $|0\rangle$ and
$|1\rangle$ {\it at the same time}. For instance, the superposition
$|0\rangle+|1\rangle$ has equal amplitudes for each basis state,
meaning that there is a 50\% probability of measuring the qubit in
$|0\rangle$ and a 50\% probability in $|1\rangle$.  Superpositions of
quantum states are the source of the interference that drives a
quantum computer; quantum algorithms attempt to manipulate the {\em
amplitude and phase} of various states so that desirable states (the
answers to the problem being solved) have a high probability of being
measured while the undesirable states (the non-answers to the problem
being solved) have a low probability of being measured.

Superposition can extend beyond single qubits and can be seen in
multi-qubit situations. Two qubits (labelled $A$ and $B$) can exist in
a quantum state such as
\begin{eqnarray}
|\psi\rangle_{AB}=|0\rangle_{A}|1\rangle_{B}-|1\rangle_{A}|0\rangle_{B}.
\end{eqnarray}
In this interesting state, if we measure the first qubit to be in the
state $|0\rangle_{A}$, then the second qubit has to be in the state
$|1\rangle_{B}$; conversely, getting the measurement result
$|1\rangle_{A}$ guarantees that we will find $|0\rangle_{B}$. The A
and B measured results are perfectly anti-correlated. This multi-qubit
superposition described above is generally given the special name {\it
entanglement} because neither qubit can be said to be in a state of
its own, independent of the other. The state cannot be factored into a
product, $|\psi\rangle_{AB}\neq |\chi\rangle_{A}|\eta\rangle_{B}$, for
any choice of basis transformation.  Many gates that act on two qubits
can change their level of entanglement, increasing or decreasing it,
depending on the gate and the initial state of the qubits.  Once a
pair of qubits are entangled, they may be separated by any distance,
and will retain their shared state.  This behavior results in the
``spooky action at a distance'' that so disturbed Einstein about
quantum theory. The maximally entangled pair of qubits are called {\em
EPR pairs} or {\em Bell pairs}, and can be used to {\em teleport}
quantum data, such that the unknown state of one qubit can be moved
from one location to another without transporting the physical carrier
of information of the qubit, consuming an EPR pair in the process
\cite{bennett:teleportation,furusawa98}.

It has been shown that entanglement between the separate quantum
computers, or nodes, of a distributed quantum system is necessary if
the system is to have the potential for exponential speedup over a
classical computer (or cluster of classical
computers)~\cite{jozsa03:entangle-speedup,love05:_type_ii_quant_algo,yepez01:_type_ii}.
At a practical level, this need for node-spanning entanglement arises
because application algorithms require gates that act on data that is
stored in separate
nodes~\cite{van-meter06:thesis,van-meter05:_distr_arith_quant_multic,yimsiriwattana04:dist-shor}.
This can be achieved by teleporting data from node to node and
performing computation locally (which we refer to as {\em teledata}),
or, alternatively, by using essentially the same techniques to execute
the equivalent of a local gate over a distance, without bringing the
two qubits together.  This technique is known as teleporting a gate
(which we refer to as {\em
telegate})~\cite{gottsman99:universal_teleport}.  We have found that,
for some application workloads and a reasonable set of physical
assumptions, it is better to teleport data than
gates~\cite{van-meter06:thesis,van-meter05:_distr_arith_quant_multic}.

\begin{figure}
\centering
\includegraphics[width=.45\textwidth]{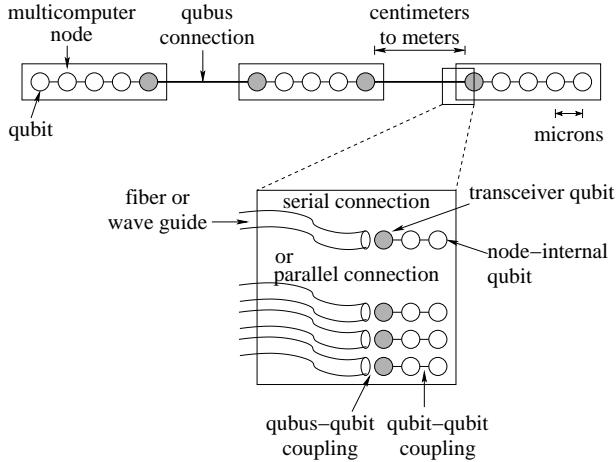}
\caption{A quantum multicomputer architecture with detail of qubus
  connections.}
\label{fig:qmc-sp-hw}
\end{figure}

Our {\em quantum multicomputer} (QMC) architecture is composed of many
small nodes, holding only a few logical qubits each, with each node
connected to two neighbors, left and right, into a line, as shown in
Figure~\ref{fig:qmc-sp-hw}.  The connections are assumed to be {\em
qubus} links, which entangle distant qubits using a strong probe laser
beam that interacts weakly with qubits connected to the bus, which we
call the {\em transceiver
qubits}~\cite{munro05:_weak,spiller05:_qubus,nemoto05:_univer}.  The
connections could also be made using single photons instead of the
qubus~\cite{cirac97:_distr_quant_comput_noisy_chann,briegel98:_quant_repeater,childress05:_ft-quant-repeater,childress2006ftq,van-Enk01091998,cirac1997qst,mancini2004eia,duan04:_scalab,duan2005rqg,van-Enk:PhysRevLett.78.4293,waks2006dit,yao05:_spin-control-prl,dur:PhysRevA.59.169,chen06:_fault,duan2001ldq,zhao06:_robust,jiang06:fast},
though the basic architecture is independent of this choice.  The
qubits may be solid-state qubits (such as quantum dots or one of
several types of Josephson junction superconducting qubits), and are
assumed to be capable of interacting with their neighbors inside a
single node or with the qubus.

The performance of any computing system must be measured with respect
to a particular workload; we have found that this configuration works
well for Shor's factoring algorithm~\cite{shor:factor}.  The most
computationally intensive portion of the algorithm is the modular
exponentiation~\cite{van-meter04:fast-modexp,beckman96:eff-net-quant-fact,vedral:quant-arith}.
This modular exponentiation is $O(n^3)$ for factoring an $n$-bit
number, both in local gate count and in teleportation operations.
Table~\ref{tab:teleport-count} shows the number of logical qubit
teleportations necessary to execute the modular exponentiation portion
of Shor's algorithm for 16, 128, and 1,024 bits.  The design choices
of the number of qubits per node and the addition algorithm to be used
are important.  The carry-lookahead adder requires ten to fifteen
times as many teleportations as the carry-ripple adders (for 16 to
1,024 bits), but may produce results faster under some circumstances;
this accounts for the range of values in
Table~\ref{tab:teleport-count}~\cite{draper04:quant-carry-lookahead,vedral:quant-arith,cuccaro04:new-quant-ripple}.
The numbers in this table are used to choose the values for the
results presented in Table~\ref{tab:teleport-code-strengths}.

\begin{table}
\caption{Number of teleportations necessary to execute the full
modular exponentiation for Shor's factoring algorithm for different
problem sizes on a quantum
multicomputer~\cite{van-meter06:thesis,van-meter05:_distr_arith_quant_multic}.}
\label{tab:teleport-count}
\centering
\begin{tabular}{|r|r|}\hline
length & teleportations ($t$) \\
\hline
16 & $14000$--$125000$\\
128 & $8\times 10^6$--$10^8$\\
1024 & $4\times 10^9$--$6\times 10^{10}$\\
\hline
\end{tabular}
\end{table}

Individual physical qubits are quite fragile and prone to errors and
deterioration over time; therefore, application-level algorithms are
generally assumed to run on logical qubits, encoded in multiple
physical qubits via {\em quantum error
correction}~\cite{shor:qecc,steane96:_qec,calderbank96:_good-qec-exists,bennett96:_5qubit-qec,laflamme96:_5qubit-qec}.
Such error codes are generally described as $[[n,k,d]]$ codes, where
$n$ is the number of lower-level qubits in a block, $k$ is the number
of logical qubits the block represents, and $(d-1)/2$ is the maximum
number of errors in the block that will not corrupt the state.  The
coding efficiency $k/n$ of quantum codes is lower than classical codes
because quantum states must be protected from errors in both value and
phase, as well as being inherently more delicate than classical
states.  Research has concentrated on $k = 1$ codes both because
simulating larger systems is difficult, and because executing logical
operations on $k = 1$ encoded states is substantially easier that $k >
1$ states.  Codes discovered early in the development of quantum
computing include the [[7,1,3]] code based on a Hamming code, the
quantum-unique [[5,1,3]] code, and the [[9,1,3]] code derived from the
simplest classical triple-redundancy protocol.  More recently, Steane
has been investigating larger known classes of classical codes for
their quantum suitability and has recommended a [[23,1,7]] code based
on a Golay code~\cite{steane02:ft-qec-overhead}, and Brun {\em et al.}
have shown how to ease some of the restrictions on the choice of code
by utilizing entanglement~\cite{ToddBrun10202006}.  In this paper, we
examine the interaction of the [[7,1,3]] and [[23,1,7]] codes with the
teleportation necessary for distributed quantum computation.  Because
the encoded states of the [[7,1,3]] code are easier to manipulate than
the states of the [[5,1,3]] code, it is generally considered more
attractive.  The [[23,1,7]] code is efficient relative to the strength
of protection provided, as we will show in Section~\ref{sec:telefail}.
Therefore, we focus on these two codes.

This paper addresses two issues relevant to the design of systems for
distributed quantum computation: the necessary strength of error
correction to provide a high probability of success of a lengthy but
finite computation when teleportation is used as described above; and
whether the quantum error correction-encoded block may be transmitted
serially or must be transmitted in parallel, which helps determine our
hardware design.  Section~\ref{sec:dist-zero} describes how
distributed logical zero states can be constructed, providing the
basis for doing error correction on logical states that span multiple
nodes.  Section~\ref{sec:dis-dat} shows the use of distributed logical
zeroes in maintaining distributed states and performing the error 
correction while the states are in motion.  The next section discusses
how different error correction codes improve the allowable
teleportation error rate, assuming that each logical qubit is
teleported in its entirety as necessary.
Section~\ref{sec:link-design} shows that serial links perform nearly
as well as parallel links, before we conclude in
Section~\ref{sec:summary}.

\section{Distributed Logical Zeroes}
\label{sec:dist-zero}

\begin{figure}
\centerline{\scalebox{0.9}{\hbox{
\input{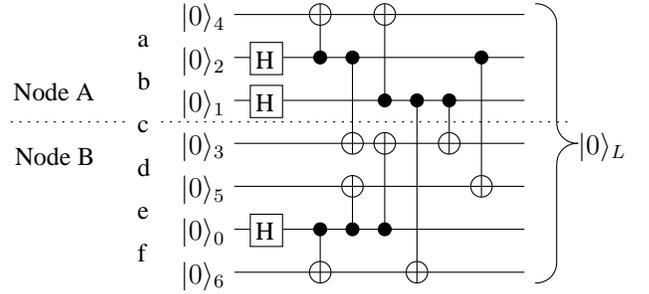}}}}
\caption{Distributed circuit to create the $|0\rangle_L$ state for the
  Steane [[7,1,3]] code.}
\label{fig:dist-713-zero}
\end{figure}

Figure~\ref{fig:dist-713-zero} shows a circuit for taking seven qubits
initialized to zero and combining them into a logical zero state
($|0\rangle_L$) for the Steane [[7,1,3]] quantum error correcting
code.  This state is used in the fault-tolerant construction of
quantum error correction and in fault-tolerant logical gates on
encoded states.  In distributed quantum computation, we may need to
perform QEC on states that span two (or more) nodes, such as during
data movement between nodes in a quantum multicomputer, or to maintain
the integrity of a static state that spans multiple nodes.  Thus, we
must find a way to either
\begin{enumerate}
\item create a distributed $|0\rangle_L$ state;
\item do parity (error syndrome) measurements using only the qubus's
  weak nonlinearity approach or single photons on four or more qubits;
  or
\item find some other way to do syndrome measurements without the
  full, distributed $|0\rangle_L$ state.
\end{enumerate}

Of these three options, we have chosen the first.  We have also
invested some effort in looking for a way to calculate the parity of
$n$ qubits using the weak nonlinearity, but all of the schemes we have
found so far for more than three qubits scale poorly in terms of
noise; Yamaguchi {\em et al.} have designed a method that works for
three qubits but not more~\cite{yamaguchi05:_weak-nonlin-qec}.  Bacon
has developed a new method for creating self-correcting memories,
using the original Shor [[9,1,3]] code, that may not require the
creation of logical zeroes; its implications for actual implementation
are exciting but still poorly
understood~\cite{bacon05:_operator-self-qec,thaker06:_cqla}.  Thus,
$|0\rangle_L$ states must be created, and this section discusses the
performance and error characteristics of the creation process.


The logical $|0\rangle_L$ can be created using the same two methods as
any other distributed quantum computation: we can directly create the
state in a distributed fashion, using teleported gates (telegate), or
we can create the state within a single node and teleport several of
the qubits to the remote node before using the state in our QEC
(teledata).  First, consider the use of teleported gates to create the
$|0\rangle_L$ state.  Figure~\ref{fig:dist-713-zero} shows that
splitting the $|0\rangle_L$ state across two nodes, as at the line
labeled ``c'', forces the execution of four teleported CNOTs,
consuming four EPR pairs; breaking at ``d'' would require only three.
In the figure, the subscripts represent the bit number in the QEC
block; the qubits have been reordered compared to the common
representation for efficiency.  Our second alternative is to teleport
portions of a locally-created $|0\rangle_L$ state.  If enough qubits
and computational resources are available at both nodes, we are free
to create the state in either location and teleport some of the
qubits; thus, the maximum number of qubits that must be teleported is
$\lfloor n/2\rfloor$, or 3 for the 7-bit Steane code.
Table~\ref{tab:713-breakpoints} shows the number of gate or data
teleportations necessary, depending on the breakdown of qubits to
nodes, showing that teledata requires the same or fewer EPR pairs, and
so is preferred.

\begin{table}
\caption[Breakpoints for the Steane {[[7,1,3]]} code]{Breakpoints
  (corresponding to Figure~\ref{fig:dist-713-zero}) and the cost of
  telegate v. teledata to create a logical zero state for the Steane
  [[7,1,3]] code, in EPR pairs consumed.  The direction qubits must be
  teleported is also shown for teledata.}
\label{tab:713-breakpoints}
\centering
\begin{tabular}{|c|c|c|}\hline
breakpoint & telegate & teledata \\
\hline
a & 2 & 1 ($B\rightarrow A$) \\
b & 3 & 2 ($B\rightarrow A$) \\
c & 4 & 3 ($B\rightarrow A$) \\
d & 3 & 3 ($A\rightarrow B$) \\
e & 3 & 2 ($A\rightarrow B$) \\
f & 2 & 1 ($A\rightarrow B$) \\
\hline
\end{tabular}
\end{table}

\section{Distributed Data}
\label{sec:dis-dat}

\subsection{Static Distributed States}

If a logical data qubit $|\psi\rangle_L$ is split between nodes A and
B in the same fashion as Figure~\ref{fig:dist-713-zero}, we will use
distributed $|0\rangle_L$ states to calculate the syndromes for the
error correction.  Each syndrome calculation consumes one
$|0\rangle_L$ state, first executing some gates to entangle it with
the logical data qubit, then measuring the zero state.  The [[7,1,3]]
code requires six syndrome measurements (three ``value'' and three
``phase'' measurements), and Steane recommends measuring each syndrome
at least twice, so each QEC cycle consumes at least a dozen logical
zero states.  With $|\psi\rangle_L$ divided at the ``d'' point, each
$|0\rangle_L$ creation requires three teleportations, for a total of
$3\times 12 = 36$ EPR pairs destroyed to execute a single, full cycle
of QEC.

The split described here allows a single logical qubit plus its QEC
ancillae, a total of fourteen physical qubits, to be split between two
nodes.  The same principles apply to states split among a larger
number of nodes, potentially allowing significantly smaller nodes to
be useful, or allowing larger logical encoding blocks to used, spread
out among small, fixed-size nodes.  More importantly for our immediate
purposes, this analysis serves as a basis for considering the movement
of logical states from node to node.

\subsection{States in Motion}

When considering the teleportation of logical qubits and their error
correction needs, two general approaches are possible:
\begin{enumerate}
\item Transfer the entire QEC block, then perform QEC locally at the
  destination; or
\item use one of the methods described above for distributed QEC {\em
  between} the teleportations of the component qubits.
\end{enumerate}

\begin{figure*}
\centerline{\hbox{
\resizebox{16cm}{!}{\input{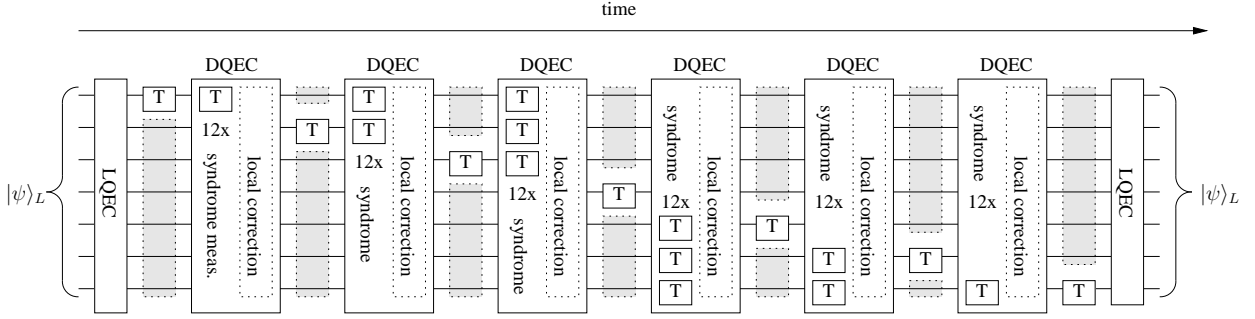}}}}
\caption{Teleporting logical state using intermediate, teledata
distributed QEC.}
\label{fig:dqec}
\end{figure*}

The analysis in Section~\ref{sec:telefail} assumes the first approach,
which is conceptually simpler; does the second approach, shown in
Figure~\ref{fig:dqec}, offer any advantages in either performance or
failure probability?  Using this approach, we attempt to reduce the
overall error probability by incrementally correcting the logical
state as it is teleported; to teleport the seven-bit state we perform
local QEC before beginning, then do distributed QEC after each of the
first six teleportations, then local QEC again after the seventh
teleportation.  Each distributed QEC (DQEC) block performs twelve
distributed syndrome measurements.  We can again choose telegate or
teledata for the $|0\rangle_L$ state creation; the figure illustrates
teledata.  Using telegate, we would need the sum of the telegate
column in Table~\ref{tab:713-breakpoints}, or $2+3+4+3+3+2=17$,
inter-node gates, for each syndrome that must be measured.  To perform
twelve measurements we consume a total of $12\times 17 = 204$ EPR
pairs.  Using teledata, we would need only $1+2+3+3+2+1 = 12$ per
syndrome, or 144 EPR pairs for the full twelve syndromes in a cycle.
The worst-case DQEC block is $3\times 12 = 36$ teleportations.
Obviously, the probability of error is higher for 36 teleportations
than for seven.  Therefore, unless someone develops a means of
measuring syndromes without using the $|0\rangle_L$ states, this
second approach does not achieve its goal of reducing the total error
probability.  Performance-wise, the penalty for doing step-wise QEC is
also stiff; we conclude that this approach is not useful, given our
current knowledge.

\section{Teleportation Failure Rates}
\label{sec:telefail}

Teleportation is composed of several phases: EPR pair creation, local
gates, measurements, and classical communication.  The EPR pairs
necessary for teleportation can be created over a fiber, interacting
with a qubit at each end via single-photon methods, or a qubus that
utilizes a strong probe beam and a weak nonlinearity, as noted above.
Until we take up the issue of link design in
Section~\ref{sec:link-design}, we will assume that local gates,
memory, and measurements are perfect, or at least much better than EPR
pair creation.  Therefore, when we talk about limits on the failure
rate of teleportation, we are really referring to the fidelity
(quality) of the EPR pair.  The quality can be improved via
purification~\cite{bennett95:_concen,cirac97:_distr_quant_comput_noisy_chann,briegel98:_quant_repeater,hartmann06},
which has a cost logarithmic in the starting fidelity; in this paper,
we will not pursue further the best way to achieve EPR pairs of the
necessary quality, though our results here may help to establish the
target fidelity for qubit purification.

The argument here falls much along the lines of the threshold argument
for quantum computation in
general~\cite{aharonov99:_threshold,preskill98:_reliab_quant_comput}.
Because we are dealing with a small number of levels of concatenation
and a finite computation, we are less interested in the threshold
itself than a specific calculation of the success probability for a
chosen arrangement.  A more detailed estimate considering all three
separate error sources in memory, local gates, and teleportation,
along the lines of Steane's
simulations~\cite{steane02:ft-qec-overhead} would differ slightly;
here we restrict ourselves to a simple analysis involving
teleportation errors only, while in later sections we will introduce
memory errors, as well.

First, let us briefly consider the failure probability assuming no
error correction on our qubits.  The probability of success of the
entire computation, then, rests on the success of {\em all} of the
individual teleportation operations.  If $t$ is the total number of
teleportations we must execute for the complete computation and $p_t$
is the probability of failure of a single teleportation, our success
probability is
\begin{equation}
p_s = (1-p_t)^t = 1 - \binom{t}{1}p_t +
\binom{t}{2}(-p_t)^2 \cdots \approx 1 - tp_t
\label{eq:p_s}
\end{equation}
for $tp_t \ll 1$.  Our failure probability grows linearly with the
number of teleportations we must execute, requiring $p_t \ll 1/t$.
Error rates of $10^{-5}$ to $10^{-11}$ are unlikely to be
experimentally achievable in the near future, so we quickly conclude
that error correction on the logical states being transferred is
necessary.

We have examined one-level QEC and two-level concatenated QEC.  We
have evaluated all of the one- and two-layer combinations of [[7,1,3]]
and [[23,1,7]].  For $p_t \ll 1$, most failures will occur in the
lowest failure mode, $((d-1)/2)+1 = (d+1)/2$ errors.  We will
approximate our total failure probability as the probability of
$(d+1)/2$ errors occurring.  The [[7,1,3]] code can restore the
correct state only when at most one component qubit has been
corrupted.  The [[23,1,7]] code can defend against three errors, so we
are interested in the probability of two and four errors,
respectively, when using these codes.


\begin{figure}
\centerline{\scalebox{0.9}{\hbox{
\input{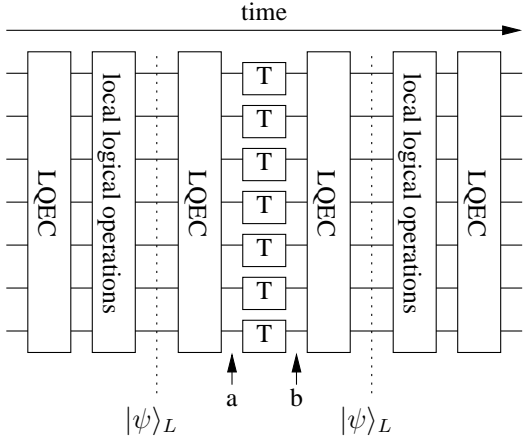}}}}
\caption{Spatially multiplexed, block-level circuit for teleporting
logical state using local QEC only, no intermediate QEC, over a
parallel interface.  The box holding a ``T'' is the teleportation
circuit.  Each line represents a qubit variable, independent of its
location, so that the teleportation operation does not explicitly show
the movement of the qubit from one node to another.  LQEC, local
quantum error correction.}
\label{fig:lqec-parallel}
\end{figure}

Transferring the seven-qubit error correction code word from one
quantum computer node to another, illustrated in
Figure~\ref{fig:lqec-parallel}, consumes seven EPR pairs.  The
probability of $m$ errors occurring is
\begin{equation}
p_e(n,m) = \binom{n}{m}(1-p_t)^{n-m}p_t^m \approx \binom{n}{m}p_t^m
\label{eq:p_e}
\end{equation}
for small $p_t$.

If $p_f$ is the failure probability of our total algorithm and $t$ is
the {\em total} number of {\em logical} qubit teleportations we use in
the computation, then
\begin{equation}
p_f = 1 - p_s = 1 - (1 - p_e)^t \approx \binom{t}{1}p_e = tp_e.
\label{eq:p_f}
\end{equation}
For this approximation to be valid, we require $tp_e \ll 1$.  For the
[[7,1,3]] code,
\begin{equation}
p_e(7,2) = \binom{7}{2} (1-p_t)^5 p_t^2 \approx 21p_t^2
\label{eq:p_e_713}
\end{equation}
is the probability of two errors occurring in our block of seven
qubits.  Two qubit errors, of course, is more than the [[7,1,3]] code
can correct.  Our probability of algorithm failure becomes
\begin{equation}
p_f \approx tp_e = 21tp_t^2.
\label{eq:p_f_713}
\end{equation}
Thus, we can say that, to have a reasonable probability of success, we
should have $p_t \ll 1/\sqrt{21t}$.  This is a significant improvement
over the case with no error correction seen above, but is still a
stringent physical condition to meet if $t$ is large.  For the
stronger [[23,1,7]] code,
\begin{equation}
p_e(23,4) = \binom{23}{4} (1-p_t)^{19} p_t^4 \approx 8855p_t^4
\label{eq:p_e_2317}
\end{equation}
implying a desired $p_t \ll 1/\sqrt[4]{8855t} = 1/9.7\sqrt[4]{t}$.

For two levels of the [[7,1,3]] code, our total encoding will consist
of seven blocks of seven qubits each, and the computation will fail
only if {\em two} or more of those blocks fail.  Of course, when using
concatenation, the two codes need not be the same.  Adapting Steane's
terminology and notation, will refer to the physical-level code as the
``inner'' code, and the code built on top of that as the ``outer''
code~\cite{steane02:ft-qec-overhead}.  [[$n^i$,$k^i$,$d^i$]] or
[[$n$,$k$,$d$]]$^i$ is the inner code, and [[$n^o$,$k^o$,$d^o$]] or
[[$n$,$k$,$d$]]$^o$ is the outer code.  Approximating the error
probability according to Equations~\ref{eq:p_e} and \ref{eq:p_f}, we
have
\begin{equation}
p_f \approx t\binom{n^o}{m^o}\left(\binom{n^i}{m^i}p_t^{m^i}\right)^{m^o}
\end{equation}
where $m^i = (d^i+1)/2$ and likewise for $m^o$.

Table~\ref{tab:teleport-code-strengths} shows the estimates for the
teleportation failure probability $p_t$ that will give us a total
algorithm failure probability of $p_f \le 0.1$.  The column titled
``scale-up'' is the number of physical qubits necessary to represent a
logical qubit.  Although [[23,1,7]]$^i$+[[7,1,3]]$^o$ and
[[7,1,3]]$^i$+[[23,1,7]]$^o$ are different, by coincidence, their
failure probabilities are almost identical.  Note that [[23,1,7]]
offers essentially the same error protection as
[[7,1,3]]$^i$+[[7,1,3]]$o$, despite using half the number of qubits
and being conceptually simpler.

From this analysis, we see that teleportation errors of $1\%$ or more
allow factoring of a 1,024-bit number on a quantum multicomputer.  In
this multicomputer, each of the 1,024 nodes contains nine logical
qubits at a scale-up of 529, for a total of almost 5,000 physical
qubits per node, when the Vedral-Barenco-Ekert (VBE) modular
exponentiation algorithm is used.  Seven of these logical qubits are
used for the VBE algorithm, and one as a buffer for each teleportation
link.  Requirements for additional ancillae used for fault tolerance
may increase the needed number of physical qubits by an amount
dependent on the speed of the underlying technology at creating
high-quality zero states and the need for local error correction.

\begin{table*}
\caption[An estimate of the necessary teleportation error rate]{An
estimate of the necessary error rate of teleportation ($p_t$) to
achieve a specific number of logical teleportations with 90\%
probability of success ($p_f = 0.1$) for the {\em entire} computation,
for different error-correction schemes.}
\label{tab:teleport-code-strengths}
\centering
\begin{tabular}{|l|c|c|c|}\hline
error-correcting code & scale-up & teleportations ($t$) & allowable
teleportation error rate $p_t$ for $p_f = 0.1$\\
\hline
(none) & 1 & $10^{5}$ & $p_t \le p_f/t = 0.1/t = 10^{-6}$ \\
& & $10^{8}$ & $ 10^{-9}$  \\
& & $10^{11}$ & $ 10^{-12}$  \\
\hline
[[7,1,3]] & 7 & $10^{5}$ & $p_t \le \sqrt{p_f/21t} = \sqrt{0.1/21t} = 2.2\times 10^{-4}$ \\
& & $10^{8}$ & $7\times 10^{-6}$ \\
& & $10^{11}$ & $2.2\times 10^{-7}$ \\
\hline
[[23,1,7]] & 23 & $10^{5}$ & $ p_t \le \sqrt[4]{p_f/8855t} = 1/17\sqrt[4]{t}\approx 3.3\times 10^{-3}$ \\
& & $10^{8}$ & $5.8\times 10^{-4}$ \\
& & $10^{11}$ & $1\times 10^{-4}$ \\
\hline
[[7,1,3]]$^i$+[[7,1,3]]$^o$ & 49 & $10^{5}$ & $p_t \le \sqrt{\sqrt{p_f/21t}/21} = 0.057/t^{1/4}
\approx 3.2\times 10^{-3}$ \\
& & $10^{8}$ &  $ 5.7\times 10^{-4}$ \\
& & $10^{11}$ &  $ 1\times 10^{-4}$ \\
\hline
[[23,1,7]]$^i$+[[7,1,3]]$^o$ & 161 & $10^{5}$ & $p_t \le \sqrt[4]{\sqrt{p_f/21t}/8855} = 0.053/t^{1/8} \approx 0.013$ \\
 & & $10^{8}$ & $5.3\times 10^{-3}$ \\
& & $10^{11}$ & $2.2\times 10^{-3}$ \\
\hline
[[7,1,3]]$^i$+[[23,1,7]]$^o$ & 161 & $10^{5}$ & $p_t \le \sqrt{\sqrt[4]{p_f/8855t}/21} = 0.053/t^{1/8} \approx 0.013$ \\
 & & $10^{8}$ & $5.3\times 10^{-3}$ \\
& & $10^{11}$ & $2.2\times 10^{-3}$ \\
\hline
[[23,1,7]]$^i$+[[23,1,7]]$^o$ & 529 & $10^{5}$ & $p_t \le \sqrt[4]{\sqrt[4]{p_f/8855t}/8855} = 0.051/t^{1/16} \approx 0.025$ \\
& & $10^{8}$ & $0.016$ \\
& & $10^{11}$ & $0.010$ \\
\hline
\end{tabular}
\end{table*}

\section{Implications for Link Design}
\label{sec:link-design}

The performance of error correction influences an important hardware
design decision: should our network links be serial or parallel?  We
can multiplex the transfer of the qubits either temporally or
spatially, as shown in Figure~\ref{fig:qmc-sp-hw}.  The figure shows
qubus fibers or wave guides coupling to one or more qubits.  In the
figure, the fiber and qubit are drawn approximately the same size, but
in reality the fiber or wave guide is likely to be many times the size
of the qubit.  Thus, these connections may require large amounts of
die space, force large qubit-qubit spacing (which affects the quality
of interaction for some types of qubits), and make high-quality
connections difficult, reducing manufacturing yield.  Each qubus
connection is therefore expensive, and minimizing their number is
desirable.  We argue that the difference in both reliability and
performance is likely to be small, assuming that the reliability of
teleportation is less than that of quantum memory and that
teleportation times are reasonable compared to the cycle time of
locally-executed QEC.

\begin{figure}
\resizebox{7.5cm}{!}{
\input{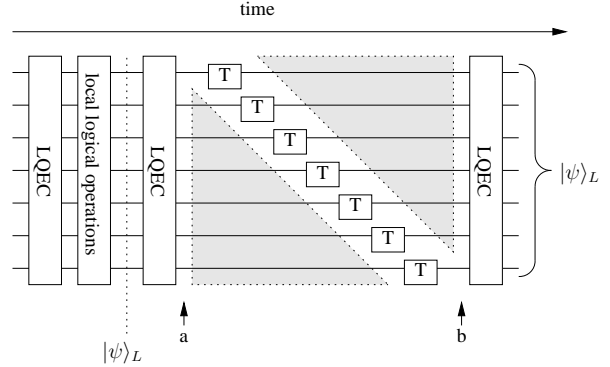}}
\caption{Temporally multiplexed, logical circuit for encoded state
transfer over a serial interface.  LQEC, local quantum error
correction.  Teleporting logical state using local QEC only, no
intermediate QEC.  The box holding a ``T'' is the teleportation
circuit.  Each line represents a qubit variable, independent of its
location, so that the teleportation operation does not explicitly show
the movement of the qubit from one node to another.}
\label{fig:lqec-serial}
\end{figure}

Figure~\ref{fig:lqec-parallel} shows a [[7,1,3]] state being
transferred in parallel and Figure~\ref{fig:lqec-serial} shows the
serial equivalent.  In these diagrams, each line represents a qubit
that is a member of a code block, essentially following the variable
rather than the storage locations; at a $T$ block, representing
teleportation, of course the qubit moves from one node to the other.
If the transfer is done serially, the wait to {\em start} the QEC
sequence is seven times as long, but the {\em total} time for transfer
plus QEC (that is, time from the start of one QEC cycle to the next,
from the first $|\psi\rangle_L$ to the point marked ``b'' in the
figures) won't grow by nearly as large a factor if local QEC requires
significant time compared to a teleportation.  Thus, we need to
determine if the increase in wait time caused by the lengthening of
the interval from the point marked ``a'' to the point marked ``b'' in
Figures~\ref{fig:lqec-parallel} and \ref{fig:lqec-serial} has an
unacceptably large impact on our overall failure rate.

The gray areas in Figure~\ref{fig:lqec-serial} indicate increased wait
time for the qubits.  They total $n(n-1)$ for an [[$n$,$k$,$d$]] QEC
code.  For the [[7,1,3]] code, each qubit spends one cycle
teleporting, and six waiting for the other teleportations.  If $p_m$
is the probability of error for a single qubit during the time to
execute a single teleportation, then the probability of no error on
one bit during that time is $(1-p_m)^6$ for the [[7,1,3]] code.  For
an [[$n$,$k$,$d$]] code, the failure probability of that qubit during
the serial transfer waiting time is $p'_m = 1 - (1-p_m)^{n-1}$.  The
probability of $m$ memory errors is
\begin{equation}
\begin{aligned}
p_M(n,m) &= \binom{n}{m}{p'_m}^m(1-p'_m)^{n-m} \\
&\approx \binom{n}{m}{p'_m}^m \\
&\approx \binom{n}{m}(n-1)p_m^m.
\end{aligned}
\label{eq:tele-mem-prob}
\end{equation}

Combining Equations~\ref{eq:tele-mem-prob} and \ref{eq:p_e}, we need
the two error sources together to generate less than $m = (d+1)/2$
errors.  We will constrain the final combined memory and teleportation
error rate $p_f$ for the serial link to be similar to the
teleportation errors for the parallel link,
\begin{equation}
p_f(n,m) = \sum_{i=0}^{m}p_M(n,i)p_e(n,m-i) \sim p_e(n,m).
\label{eq:tele-mem-comp}
\end{equation}

For the error codes we are considering, [[7,1,3]] and [[23,1,7]],
numeric evaluation for $p_m = p_t/10(n-1)$ gives 25\% and 50\%
increase in failure probability, respectively, compared to the $p_m =
0$ (perfect memory) case.  Thus, we can say, very roughly, that a
memory failure probability two orders of magnitude less than the
failure probability of the teleportation operation will mean that the
choice of serial or parallel buses has minimal impact on the overall
system error rate.

Although this section has focused on reliability rather than
performance, the choice of serial or parallel links also affects
performance.  It is easy to see that choosing a serial link does not
result in a factor of $n$ degradation in system performance when QEC
is taken into account.  Let $t_t$ be our teleportation time, and
$t_{LQEC}$ be the time to perform local error correction.  $t_t$ is
related to the detector time for measuring the probe beam on the
long-distance links, while $t_{LQEC}$ is related to the local qubit
measurement time.

If $nt_t \ll t_{LQEC}$, then in accordance with Amdahl's Law the
choice also has minimal impact on our overall
performance~\cite{amdahl67}.  Moreover, for Shor's algorithm on the
quantum multicomputer, we have shown that breaking down the
teleportation operation into its component phases of EPR pair creation
and the measurement and classical operations allows application-level
performance to be relatively independent of the quantum link operation
time~\cite{van-meter05:_distr_arith_quant_multic}.  Therefore, we
recommend using serial links.

\section{Summary}
\label{sec:summary}

This paper has tackled two important issues in the design of
distributed quantum computing systems, both centering around the need
to correct errors that occur during teleportation, analyzed in the
context of a long but finite computation such as Shor's factoring
algorithm.  We have shown that a relatively high failure rate for
teleportation is tolerable, and that using serial links rather than
parallel has only a modest impact on the probability of
failure and the performance of the computation.

The results in Table~\ref{tab:teleport-code-strengths} show that a
teleportation error rate (related to the EPR pair infidelity) of
$>1\%$ will allow computations as large as the factoring of a
1,024-bit number to proceed with a high probability of success.  This
estimate is for a data encoding of [[23,1,7]]$^i$+[[23,1,7]]$^o$ on
the link.  Our analysis supports Steane's recommendation of the
[[23,1,7]] code.  Replacing one level with the [[7,1,3]] code still
allows an error rate of one part in a thousand or better, with a
noticeable savings in storage requirements.  Of course, we do not have
to compute or store data using the same encoded states that we use
during data transport, as noted by Thaker {\em et
al.}~\cite{thaker06:_cqla}.  In this paper, for simplicity, we have
assumed that the system uses only a single choice of encoding.

We have argued that the difference in both performance and reliability
between serial and parallel network links will be small for a
reasonable set of assumptions.  A memory error rate in the time it
takes to perform a teleportation at least two orders of magnitude
better than the teleportation failure rate results in a $25-50\%$
increase in the computation failure rate, an increase we consider
acceptable in exchange for the benefits of serial links.  Serial links
will dramatically simplify our hardware design by reducing the number
of required transceiver qubits in each node, and eliminating concerns
such as jitter and skew between pairs of conductors or wave guides.
Moreover, if we do choose to have multiple transceiver qubits in each
node, system performance on some workloads may be boosted more by
creating a richer node-to-node interconnect topology than by creating
parallel channels between pairs of nodes in a simpler topology.

\section*{Acknowledgments}

The authors thank MEXT and QAP for partial support for this research.
We thank Kohei M. Itoh for technical help.


\end{document}